# Sparsity-regularized coded ptychography for robust and efficient lensless microscopy on a chip

Ninghe Liu[1*], Qianhao Zhao[2], Guoan Zheng[2,3*]

[1]Department of Electrical Engineering, California Institute of Technology, Pasadena, CA 91125, USA
[2]Department of Biomedical Engineering, University of Connecticut, Storrs, CT 06269, USA
[3]Center for Biomedical and Bioengineering Innovation, University of Connecticut, Storrs, CT 06269, USA
[*]Correspondence: N. L. (nliu@caltech.edu) or G. Z. (guoan.zheng@uconn.edu)

**Abstract:** Coded ptychography has emerged as a powerful technique for high-throughput, high-resolution lensless imaging. However, the trade-off between acquisition speed and image quality remains a significant challenge. To address this, we introduce a novel sparsity-regularized approach to coded ptychography that dramatically reduces the number of required measurements while maintaining high reconstruction quality. The reported approach, termed the ptychographic proximal total-variation (PPTV) solver, formulates the reconstruction task as a total variation regularized optimization problem. Unlike previous implementations that rely on specialized hardware or illumination schemes, PPTV integrates seamlessly into existing coded ptychography setups. Through comprehensive numerical simulations, we demonstrate that PPTV-driven coded ptychography can produce accurate reconstructions with as few as eight intensity measurements, a significant reduction compared to conventional methods. Convergence analysis confirms the robustness and stability of the PPTV algorithm. Experimental results from our optical prototype, featuring a disorder-engineered surface for wavefront modulation, validate PPTV's ability to achieve high-throughput, high-resolution imaging with a substantially reduced measurement burden. By enabling high-quality reconstructions from fewer measurements, PPTV paves the way for more compact, efficient, and cost-effective lensless microscopy systems on a chip, with potential applications in digital pathology, endoscopy, point-of-care diagnostics, and high-content screening.

**Introduction**
In optical imaging, the phase of light is crucial for understanding its behavior, yet detectors can only capture light intensity, making phase retrieval a challenging, ill-posed problem. Ptychography emerged as a powerful technique to address this limitation, initially developed for electron crystallography [1] and later adapted for optical applications. By acquiring multiple diverse diffraction patterns through systematic scanning of a coherent probe beam across a specimen, ptychography enhances the retrieval of both phase and amplitude information. Modern ptychography systems employ iterative phase retrieval frameworks, applying constraints in real and reciprocal space to refine estimates [2, 3]. The technique's success relies on multiple overlapping areas of illumination, which mitigates inherent ambiguities present in single-measurement phase retrieval. In the past decades, ptychography has rapidly evolved and diversified into various forms and applications [4-18].

The principle of multiple overlapping illumination areas is fundamental to successful ptychographic reconstruction. In traditional single-measurement phase retrieval, multiple potential objects with varying amplitude and phase could mathematically yield identical intensity patterns at the sensor plane. Ptychography effectively addresses these ambiguities by imposing diverse probe beams on the same objects, particularly within overlapping regions, thus significantly improving the likelihood of successful recovery. However, achieving high accuracy and resolution in ptychographic reconstruction often requires hundreds or even thousands of measurements. This approach results in a time-consuming and equipment-intensive process, creating a trade-off between measurement quantity and resultant image quality. The need for numerous measurements poses a significant challenge in practical applications, particularly in scenarios requiring rapid or real-time imaging.

To address the challenge of extensive measurements and unlock the full potential of limited data sets, we draw inspiration from compressive phase retrieval techniques [19, 20]. These methods promise to push phase recovery beyond conventional information-theoretic limits by extensively exploring signal priors. In single-shot phase retrieval and holography, sparsity priors in spatial [21-23], gradient [24-30], wavelet [31], and other forms [32, 33] have proven to be effective regularizers. This success provides a theoretical foundation for introducing sparsity into wide-field, high-resolution ptychographic reconstruction.

Our work focuses on leveraging gradient-domain sparsity to accelerate ptychographic acquisition and reconstruction. We formulate the reconstruction problem as an objective function that combines a fidelity term, representing the disparity between the recovered wavefield and intensity measurements, with a total variation (TV) regularization term. The piece-wise smoothness property of the TV model and its computationally efficient algorithms [30, 34, 35] enable our proposed reconstruction model to represent multidimensional features while ensuring generalizability to various sample scenarios. We introduce a ptychographic proximal TV-solver (PPTV) and integrate this optimization framework into the coded ptychography system without hardware modifications. Both numerical simulations and experimental results demonstrate that the PPTV-driven approach substantially reduces the required number of intensity measurements while maintaining high-quality results. The development of PPTV represents a contribution towards alleviating the trade-off between high-quality ptychographic imaging and the resource-intensive demand for numerous measurements. This advancement enhances the feasibility of more streamlined and efficient ptychographic microscopy applications, potentially impacting fields such as digital pathology, point-of-care diagnostics, and high-content screening.

**Methods**
**Preliminary illustrations on notations and forward model.** In the following sections, the notations are defined as follows: The symbol $|\cdot|_p$, $(\cdot)^T$, $(\cdot)^*$, and $(\cdot)^H$ denotes the *p*-norm, transpose, conjugate, and conjugate transpose (Hermitian) operators, respectively. $\odot$ is the element-wise (Hadamard) multiplication operator for vectors, while $|\cdot|$, $|\cdot|^2$, $\sqrt{\cdot}$, and $\cdot/\cdot$ should also be interpreted as element-wise operators when applied to vectors. $\text{diag}(\cdot)$ puts the entries of a vector into the diagonal of a matrix. For the convenience of illustration, we follow the convention of linear algebra and vectorize the two-dimensional wavefield as a one-dimensional vector representation. It is important to notice that vectorization is just for simplifying the notation and is not required for computation in practice.

In the forward model, the object $O$ is first propagated by distance $d_1$ to the coded surface (CS) plane, where it is modulated by the spatially shifted coded sensor. On account of the relativity of displacement, the exit wave $\Psi_{cs}^k$ downstream the object for the *k*-th scan position can be expressed as:

$$\Psi_{cs}^k = (H_k P_{d_1} O) \odot \phi_{cs} \tag{1}$$

where $P_{d_1}$ and $\phi_{cs}$ represent the free space propagation for distance $d_1$ and the complex transmittance matrix of the CS, respectively. $H_k$ stands for the spatial translation operation with respect to the *k*-th scan position. For the displacement $(x_k, y_k)$ caused by the *k*-th translation, noticing that translation in the spatial domain is equivalent to phase shift in the Fourier domain, $H_k$ can be written as:

$$H_k = \mathcal{F}^{-1}\{e^{j2\pi(f_x x_k + f_y y_k)} \odot \mathcal{F}\} \tag{2}$$

$\mathcal{F}$ and $\mathcal{F}^{-1}$ denote the counterpart of 2D Fourier transform and inverse 2D Fourier transform in the expression of vectorized matrix and $f_x$ and $f_y$ are the corresponding frequencies after the Fourier transform operation. We denote $\Psi_u = P_{d_1}O$ and $\Psi_{us}^k = H_k\Psi_u$ for the convenience of the following illustration. The exit wave $\Psi_{cs}^k$ is then propagated to the sensor plane, where it is down-sampled by image sensor pixels and recorded as intensity patterns. The intensity measurements $y_i$ with respect to the $k$-th scan position can be expressed as:

$$y_k = S|P_{d_2}\Psi_{cs}^k|^2 \tag{3}$$

where $P_{d_2}$ and $S$ denote the free space propagation for distance $d_2$ and the downsampling matrix of the image sensor, respectively. With sensor downsampling, Eq. (3) can be rewritten as:

$$y_k = B|P_{d_2}\Psi_{cs}^k|^2 + b \tag{4}$$

where $B$ is a diagonally-dominant positive-definite matrix and $b$ is a positive offset vector. For the simplest case we consider $B$ as a discretized 0-1 matrix that fully collects all the energy on its pixels and $b$ equals zero because, for our system depicted in Fig. 1(a), the whole image sensor is covered by the diffused laser.

**Generalized reconstruction framework for ptychography.** Based on the coded ptychography system depicted in Fig. 1(a), We formulate the ptychography reconstruction as a constrained optimization problem, which can also be generalized to other conventional ptychography cases where the coded surface should be replaced by a probe. Generally, the reconstruction process of PPTV can be divided into three subtasks, as shown in Fig. 1(b).

Step 1: Optimization at the sensor plane. Search for the optimal wave field $\Psi_k^{opt}$ that minimizes the distance between the intensity measurements $I_k$ and the wave field intensity $|\Psi_s^k|^2$ given by the forward model, as is shown in Fig. 1(b1).

$$(\Psi_s^k)^{opt} = \text{argmin}\, d(I_k, |\Psi_s^k|^2) \tag{5}$$

$$d\left(I_k, |\Psi_s^k|^2\right) = F(\Psi_s^k) = \frac{1}{2}\left\|\sqrt{S|\Psi_s^k|^2} - \sqrt{I_k}\right\|_2^2 \tag{6}$$

In our work, we choose the amplitude-based fidelity term as the distance metric because of its mathematical similarity to the Poisson noise model [11, 36], which is proven to be an accurate model for noise statistics description. The detailed proof substantiating this choice is provided in the Supporting Information, Section 1. To optimize the fidelity term, we employ the gradient descent method, actively seeking to determine the optimal step size. The Wirtinger gradient is given by:

$$\nabla F(\Psi_s^k) = \frac{1}{2}\text{diag}(\Psi_s^k)S^T\left(1 - \frac{\sqrt{I_k}}{\sqrt{S|\Psi_s^k|^2}}\right) \tag{7}$$

Once the locally optimal update direction is derived, we can perform the gradient descent operation:

$$(\Psi_s^k)' = \Psi_s^k - \beta_k \nabla F(\Psi_s^k) \tag{8}$$

where $\beta_k$ is the step size for the descent direction. Notice that for specific update direction, the fidelity term can be seen as a convex function with respect to $\beta_k$, hence the optimal step size can be calculated by finding the extreme of $F(\beta_k)$:

$$\partial F(\beta_k) / \partial \beta_k = 0 \tag{9}$$

Solving Eq. (8) yields $\beta_k = 2$, thus we derive the optimal wave field for $k$-th scan position to be

$$(\Psi_s^k)^{opt} = \Psi_s^k \odot S^T\left(\frac{\sqrt{I_k}}{\sqrt{S|\Psi_s^k|^2}}\right) \tag{10}$$

which is equivalent to the well-known modulus constraint projection method [2].

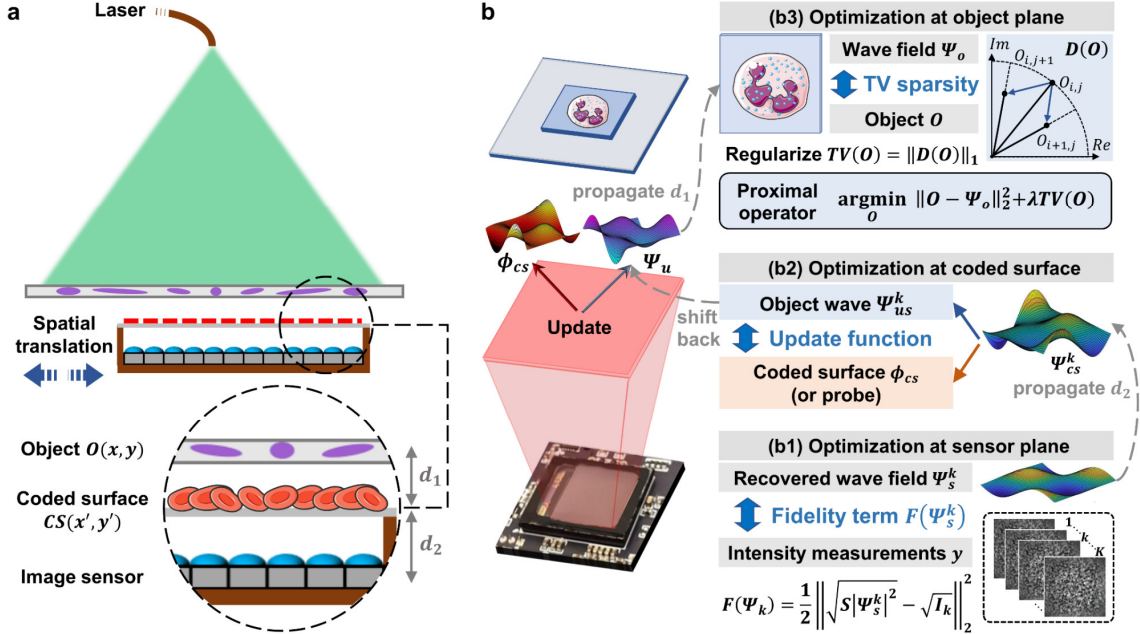

**Figure 1. Coded ptychography system setup and generalized PPTV reconstruction framework.** (a) Schematic of the coded ptychography device. A collimated laser beam illuminates the object. The object is placed above a coded surface (e.g., blood-coated sensor), which modulates the transmitted wavefront. The sensor translates laterally to acquire multiple diffraction patterns. (b) Generalized PPTV reconstruction framework, illustrating the three key optimization steps: (b1) Optimization at the sensor plane: Enforcing fidelity between recovered wavefield and intensity measurements. (b2) Optimization at the coded surface: Updating the object wave and coded surface (or probe) profiles. (b3) Optimization at the object plane: Applying TV sparsity regularization to the object. The framework integrates wavefield propagation between planes and incorporates prior information to enhance reconstruction quality while minimizing required measurements. This approach enables efficient, high-quality ptychographic imaging with potential for compact system designs.

Step 2: Optimization at the coded surface plane. Apply a proper update function to the recovered wave field $\Psi_{cs}$ at the coded surface plane, as is shown in Fig. 1(b2). The uniqueness of ptychography is that the complex transmission matrix of the coded surface (or probe) is not accurately known, so a proper update needs to be conducted on the back-propagated wave field at the coded surface plane $\Psi_{cs}^k$ to recover the correct coded surface $\phi_{cs}$ and object-related wavefield $\Psi_u$. An error metric is introduced to evaluate the update process [37]:

$$E = \left\| \Psi_{cs}^k - (H_k \Psi_u) \odot \phi_{cs} \right\|_2^2 \tag{11}$$

where $\Psi_{cs}^k = P_{d_2}^{\mathrm{H}}(\Psi_s^k)^{opt}$. Again we calculate the Wirtinger gradient with respect to $\Psi_u$ and $\phi_{cs}$ and take the gradient descent method to optimize $E$:

$$\nabla E_\Psi(\Psi_u) = H_k^{\mathrm{H}} \left( (\phi_{cs})^* \odot \left( \Psi_{us}^k \odot \phi_{cs} - \Psi_{cs}^k \right) \right) \tag{12}$$

$$\nabla E_\phi(\phi_{cs}) = \left( \Psi_{us}^k \right)^* \odot \left( \Psi_{us}^k \odot \phi_{cs} - \Psi_{cs}^k \right) \tag{13}$$

$$\Psi_u' = \Psi_u - \beta_u^k \nabla E_\Psi(\Psi_u) \tag{14}$$

$$\phi_{cs}' = \phi_{cs} - \beta_{cs}^k \nabla E_\phi(\phi_{cs}) \tag{15}$$

where $\Psi_{us}^k = H_k \Psi_u$ denotes the shifted object wave. Ref. [37] mentions that setting $\beta_u^k = \alpha_1/|\phi_{cs}|_{max}^2$ and $\beta_{cs}^k = \alpha_2/|\Psi_{us}^k|_{max}^2$ gives the classic ePIE update function for object and probe update [4], while Ref. [11]

takes another approach by neglecting second-order terms and solving the step size with the least-square (LSQ) method. We make a slight modification to the equation brought up in Ref. [11] and give the resulting LSQ matrix:

$$\begin{pmatrix} \|\Delta\Psi_{us}^k \odot \phi_{cs}\|_2^2 + \alpha & Re\{(\Delta\Psi_{us}^k \odot \phi_{cs})^H(\Delta\phi_{cs} \odot \Psi_{us}^k)\} \\ Re\{(\Delta\Psi_{us}^k \odot \phi_{cs})^H(\Delta\phi_{cs} \odot \Psi_{us}^k)\} & \|\Delta\phi_{cs} \odot \Psi_{us}^k\|_2^2 + \alpha \end{pmatrix} \begin{pmatrix} \beta_u^k \\ \beta_{cs}^k \end{pmatrix}$$
$$= \begin{pmatrix} Re\{(\Delta\Psi_{us}^k \odot \phi_{cs})^H \mathcal{X}\} \\ Re\{(\Delta\phi_{cs} \odot \Psi_{us}^k)^H \mathcal{X}\} \end{pmatrix} \quad \beta_u^k, \beta_{cs}^k \in \mathbb{R} \quad (16)$$

where $\mathcal{X} = \Psi_{cs}^k - \Psi_{us}^k \odot \phi_{cs}$ represents the residual term and $\alpha$ is a small regularization term to prevent numerical instability. $\Delta\Psi_{us}^k = -H_k \nabla E_\Psi(\Psi_u^k)$ and $\Delta\phi_{cs} = -\nabla E_\phi(\phi_{cs})$ are the update direction for shifted object wave and coded surface. $Re$ denotes the real part of a complex number, and $\mathbb{R}$ is the real number constraint that step size ought to satisfy. Derivation of this is also provided in our Supporting Information, Section 2. In PPTV we simply choose the parameters that lead to ePIE update function in consideration of computational cost, while a comparison between PPTV and this LSQ-ML method is also given in the following illustration.

Step 3: Optimization at the object plane. Use prior information to further optimize the recovered wave field at the object plane, as is shown in Fig. 1(b3). So far, all information for the recovery process has been derived from the intensity measurements, often requiring a substantial number of measurements if no additional knowledge is available. However, in the context of PPTV, we transcend the limitations of relying solely on posterior information and incorporate TV regularization into our imaging model. This advancement allows us to inherently minimize the total variation of an image, equating to enhanced gradient sparsity—a universally applicable prior for natural images. The generalizability of PPTV is further validated through a series of numerical simulations and optical experiments with different biological samples. We consider the discrete anisotropic complex TV semi-norm depicted in Fig. 1(b3):

$$TV(O) = \|D(O)\|_1 = \sum_{i=1}^{N_1-1}\sum_{j=1}^{N_2}|O_{i+1,j}^m - O_{i,j}^m| + \sum_{i=1}^{N_1}\sum_{j=1}^{N_2-1}|O_{i,j+1}^m - O_{i,j}^m| \quad (17)$$

where $O^m \in \mathbb{C}^{N_1 \times N_2}$ with $n = N_1 \times N_2$ denotes the non-vectorized two-dimensional image. Operator $D$ computes the spatial finite difference for the vectorized object $O$ along both directions. Given the back propagated wave field $\Psi_o$, the third step of optimization can be formulated as a proximal operator:

$$prox_v = \arg\min_O \frac{1}{2}\|O - \Psi_o\|_2^2 + \lambda TV(O) \quad (18)$$

The purpose of this operator is to minimize the TV term while maintaining the proximity of our final solution to $\Psi_o$, which is derived from the preceding two optimization steps. The parameter $\lambda$ can serve as a regularization factor to control the permissible deviation from $\Psi_o$. This operator coincides with solving the TV-regularized image denoising problem [30, 34, 35, 38, 39], and the optimal solution $O^{opt}$ can be solved via the dual problem:

$$\min_{w \in S}\left\{G(w) \equiv \|\Psi_o - D^H w\|_2^2\right\} \quad (19)$$
$$W = \{w_i \mid \|w_i\|_\infty \leq \lambda\} \quad (20)$$
$$O^{opt} = \Psi_o - D^H w^{opt} \quad (21)$$

where $w \in \mathbb{C}^{2n}$ denotes the dual variable which is confined to convex set $W$, the final solution for the object $O^{opt}$ can be represented by the optimal solution for the dual variable $w^{opt}$. Proof of this is given in our Supporting Information, Section 3. Although Eq. (18) looks like a least square problem, it cannot be directly solved because the complex finite difference operator $D$ cannot be represented explicitly in a matrix form, hence the pseudo-inverse of $DD^H$ could not be derived. Alternatively, we use the accelerated gradient projection algorithm to solve this problem in a more computation-effective way. By calculating the Wirtinger gradient of $G(w)$, we iteratively perform gradient descent and project the updated solution onto set $W$ using the projection operator $P_w(\cdot)$. The step size $\eta$ for this gradient descent update is set as $1/8$ based on the convergence analysis presented in Supporting Information, Section 3.

$$\nabla_w G(w) = -D(\Psi_o - D^H w) \tag{22}$$

**Chart 1**. Solver for TV-regularized subproblem

**Input:** wave field $\Psi_o$, step size $\eta$, $\varepsilon_t = t/(t+3)$, and subiteration number $T$

1. $w^{(0)} = 0, z^{(0)} = w^{(0)}$
2. **For** $t = 1, 2, \ldots, T$ **do**
3. $w^{(t)} = P_w\left(z^{(t-1)} - \eta \nabla_z G(z^{(t-1)})\right)$      Gradient projection
4. $z^{(t)} = w^{(t)} + \varepsilon_t (w^{(t)} - w^{(t-1)})$      Nesterov's extrapolation
5. **End for**
6. $O^{opt} = \Psi_o - D^H w^{opt}$

**Objective modeling and the PPTV algorithm.** After step-by-step derivation of the PPTV algorithm, we present a unified objective modeling that integrates all three subtasks optimization, which can be expressed as:

$$\min_{O,CS} \sum_k \frac{1}{2K} \left\| A_{CS}^k O - \sqrt{I_k} \right\|_2^2 + R(O) \tag{23}$$

where $A_{CS}^k$ is the discussed forward model with respect to the coded surface $\phi_{cs}$ and the $k$-th scan position $(x_k, y_k)$, $I_k$ is the corresponding measurement intensity, and $R$ is the regularization term that represents sparsity prior. Optimization at the sensor plane and the coded surface corresponds to minimizing the first fidelity term, and optimization at the object plane corresponds to minimizing the second regularization term, which equals to $\lambda TV(O)$ in our case. We furthermore improve the algorithm's robustness and convergence speed by mini-batch optimization and Nesterov's extrapolation, and give the PPTV algorithm, as is shown in Fig. 2.

## Results

**Numerical simulations and analysis.** The advantages of the presented method are demonstrated in realistic numerical simulations because it allows us to avoid potential errors in sample positions and instability of illumination. Also, in numerical simulation reconstruction quality can be completely evaluated and analyzed quantitatively. Firstly, we assume the complex transmittance matrix of the coded surface is already known and generate the coded surface as a randomly initialized complex matrix. We compare the reconstructions of ePIE, LSQ-ML method, and PPTV in numerical simulations, as is shown in Fig. 3. The ground truth for ptychographic reconstruction is a complex wave field whose amplitude is the "street" image and phase the "peppers" image, as are shown in Fig. 3(a1) and Fig. 3(a2). A homogeneous intensity of 0.2 was added to simulate the background. In this ideal case, the coded surface is not updated, which corresponds to $j_{cs} > J_0$ in the workflow in Fig. 2. The amplitude and phase map of it are shown in Fig. 3(b1) and Fig. 3(b2). To test the performance of PPTV in case of low measurement number, we only take 8 measurements corresponding to different spatial translation locations shown in Fig. 3(c). The original image size is 1024×1024, while the sensor resolution is set to be 256×256, which means the super-resolution (SR) ratio equals 4 in our case. The iteration number is chosen so that total runtime cost of the three methods is close to each other while convergence is also guaranteed. The relative root mean square error (RMSE) is calculated between the reconstructions and the ground truth to quantitatively represent the reconstruction quality. From the results given in Fig. 3(d1-d3), we can see that in similar amount of time, the reconstruction of PPTV has the highest accuracy in overall as well as in detail: The RMSE error is reduced by an order of magnitude compared to the other two methods, and the zoomed-in amplitude and phase show excellent noise-free details.

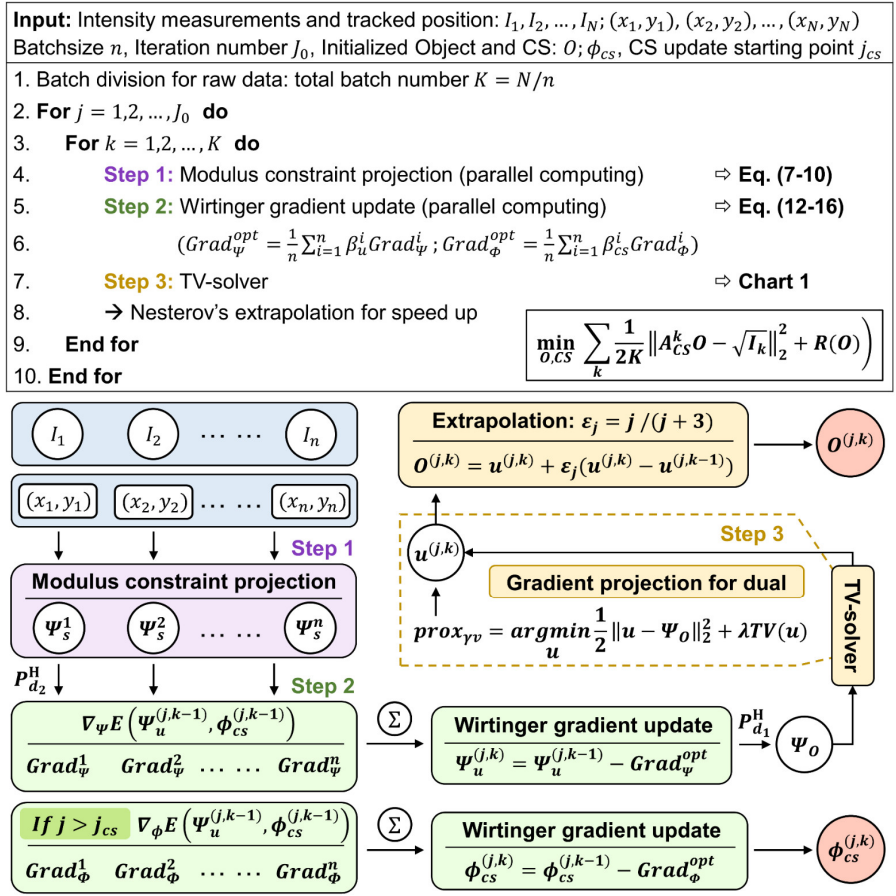

**Figure 2. Algorithm workflow of PPTV**. The three steps of optimization are integrated into an iterative optimization framework. In the practical operation of PPTV, step 1 and step 2 are performed in parallel computing. The coded surface is often not updated from the very beginning, and we denote the update starting point as $j_{cs}$.

To verify the effectiveness of the objective modeling in Eq. (23), we further analyze the convergence behavior of ePIE, LSQ-ML, and PPTV against iterations for 8-measurements numerical simulations, as is shown in Fig. 4. Specifically, we choose the fidelity error in Eq. (6) that represents the similarity between diffraction patterns and intensity measurements at the sensor plane, the objective function in Eq. (23) that embodies the essence of PPTV method, and the RMSE error that can be seen as a golden standard for reconstruction quality evaluation. For ePIE and LSQ-ML methods, the fidelity error reaches a low value and continues to decrease with each iteration in Fig. 4(a). However, the objective value has already converged and remains static after approximately 20 iterations in Fig. 4(b), a trend mirrored by RMSE in Fig. 4(c). In contrast, the PPTV approach exhibits a higher fidelity error compared to ePIE and LSQ-ML. Nevertheless, both the objective value and RMSE error for PPTV continue to decrease significantly, ultimately reaching a much lower level than the other two methods. This observation emphasizes the superior characterization of the reconstruction process by the PPTV objective modeling, which goes far beyond the fidelity error. Additionally, it is worth noting that the LSQ-ML method achieves a fidelity error lower than ePIE, yet its RMSE error is higher. This discrepancy can be attributed to the inherent ambiguities that arise when working with fewer intensity measurements, highlighting the limitations of relying solely on the fidelity error for modeling the reconstruction process and emphasizing the necessity of introducing sparsity priors, as exemplified by our PPTV approach.

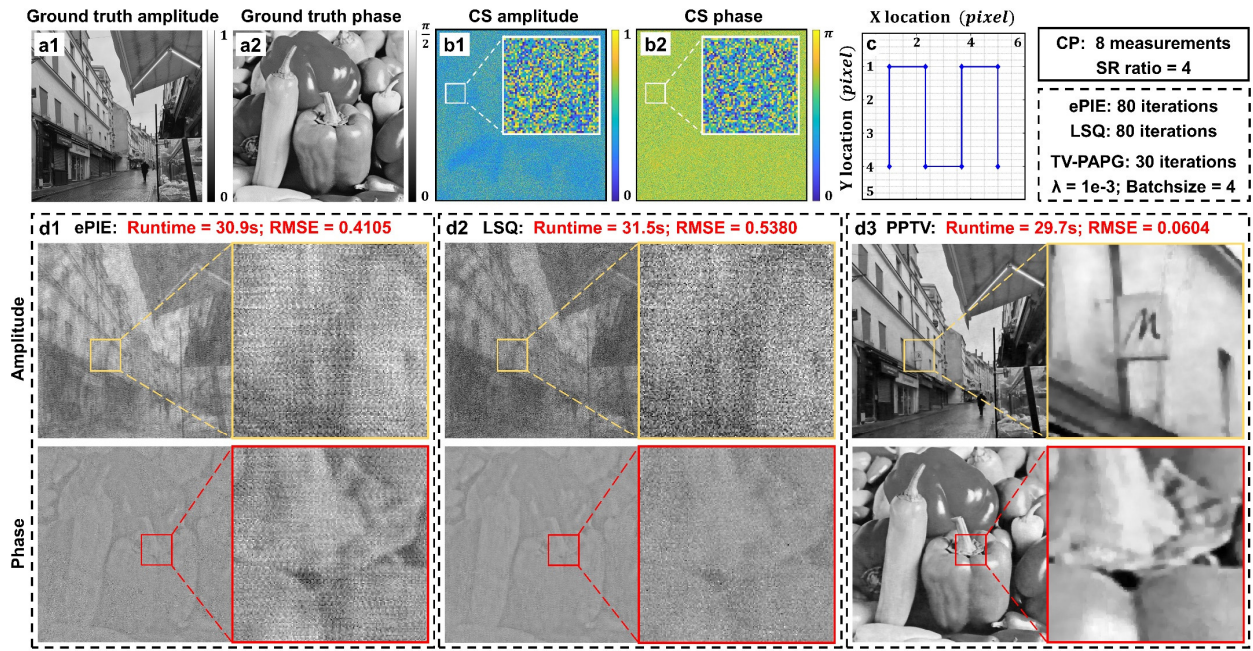

**Figure 3. Simulation parameters setup and comparison of numerical reconstructions among ePIE, LSQ-ML, and PPTV.** (a1-a2) Ground truth amplitude and phase of object. (b1-b2) Ground-truth amplitude and phase of simulated coded surface. (c) Simulated translation positions. (d1-d3) Reconstructions of objects using the ePIE, LSQ-ML, and PPTV methods. Insets show magnified regions for detail comparison. The table in the upper right corner lists key simulation parameters

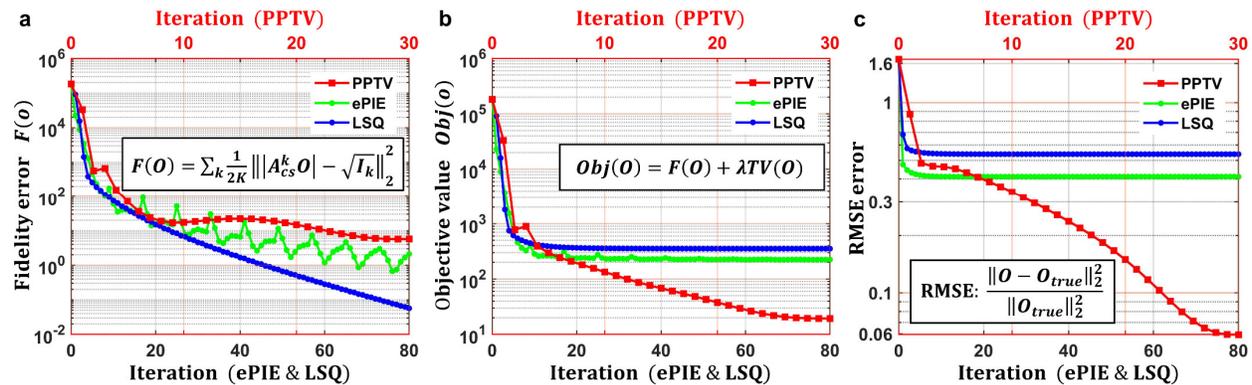

**Figure 4. Convergence analysis of ePIE, LSQ-ML, and PPTV algorithms for 8-measurement numerical simulations.** (a) Fidelity error plotted against iterations for each algorithm. The fidelity term equation is shown in the inset. (b) Objective function value versus iterations. The objective function combines the fidelity term and TV regularization, as shown in the inset. (c) RMSE plotted against iterations, with the RMSE calculation formula provided in the inset. For all plots, the x-axis shows iterations for ePIE and LSQ on the bottom (0-80) and for PPTV on the top (0-30). PPTV demonstrates faster convergence and lower final error compared to ePIE and LSQ, particularly in terms of objective value and RMSE.

In the practical application of coded ptychography and conventional ptychography, the complex transmittance matrix of coded surface (or probe) is not accurately known. To simulate this common
8

situation, we add a rather high level of Gaussian noise to the amplitude and phase of the coded surface for initialization and perform a bi-directional reconstruction on both object and coded surface:

$$\phi_{cs}^{ini} = \mathcal{P}_I\left(\left(|\phi_{cs}| + \mathcal{N}(0, \sigma_{amp})\right)\exp\left(j(angle(\phi_{cs}) + \sigma_{ang})\right)\right) \quad (24)$$

$$I = \{\phi_i \mid \|\phi_i\|_\infty \leq 1\} \quad (25)$$

where $\mathcal{N}(\mu, \sigma)$ is the normal distribution with mean and variance to be $\mu$ and $\sigma$, $angle(\cdot)$ takes the angle of a complex vector, and $\mathcal{P}_I(\cdot)$ denotes the projection operator onto the constraint $I$ and guarantees the modulus of $\phi_{cs}^{ini}$ does not exceed one. Amplitude and phase map of the disturbed coded surface are shown in Fig. 5(a1) and Fig. 5(a2). The amplitude and phase map of the complex residuals between the initialization and the ground-truth coded surface are shown in Fig. 5(b1) and Fig. 5(b2). Besides the iterations performed in the former cases, we take another same number of iterations for bi-directional update, which corresponds to $j_{cs} = J_0/2$ in the workflow in Fig. 2. Reconstruction results of the three methods are compared in Fig. 5(d1-d3), and the RMSE convergence curve is given in Fig. 5(c). Although the RMSE of PPTV reconstruction is slightly higher than the ideal case, the image quality still far exceeds ePIE and LSQ-ML reconstructions, with detailed information fully recovered, such as the letter "M" in the amplitude map and the stalk of peppers in the phase map. Furthermore, the RMSE convergence curve of PPTV exhibits a clear turn point at iteration 30. This inflection signifies that the updates applied to the object and coded surface continue to lead the reconstruction in the right direction, indicating ongoing improvement. In contrast, for ePIE and LSQ-ML methods, no apparent enhancements are evident after the initiation of coded surface updates, which demonstrates the effectiveness of PPTV in guiding the bi-directional reconstruction of both object and coded surface. We further analyze the influence of regularization parameter λ in PPTV's reconstruction, and verify that choosing $\lambda = 1 \times 10^{-3}$ achieves the best performance in our numerical setup, which is provided in the Supporting Information, Section 4.

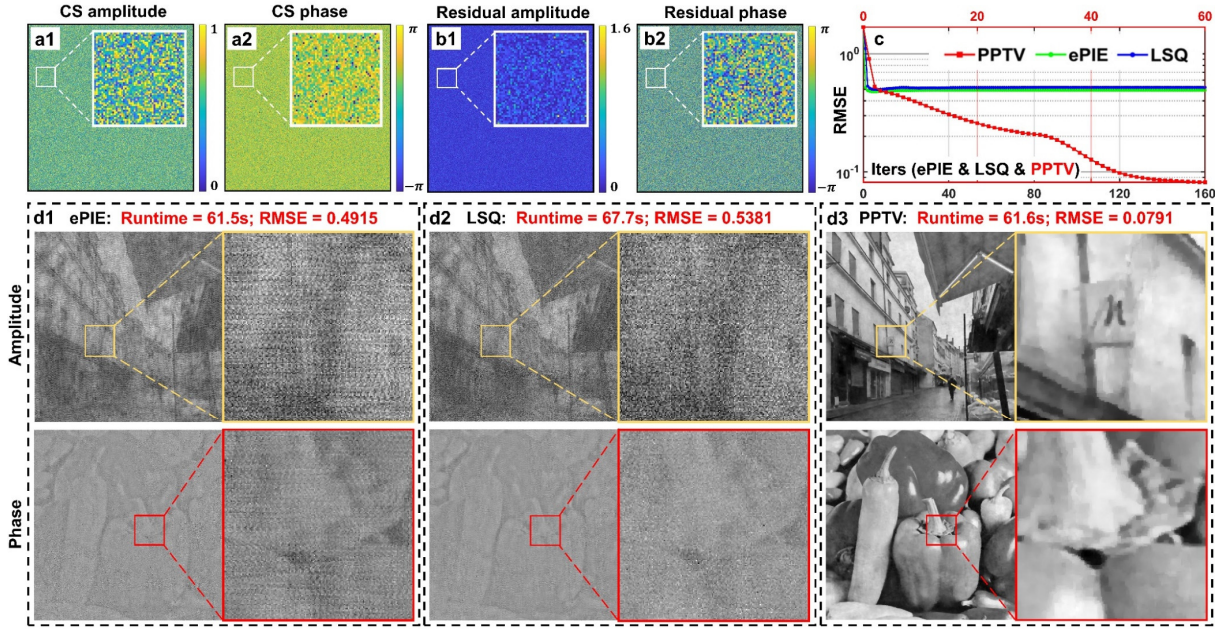

**Figure 5. Coded surface initialization, convergence analysis, and comparison of numerical reconstructions using ePIE, LSQ-ML, and PPTV methods.** (a1-a2) Amplitude and phase maps of the initialized coded surface. (b1-b2) Amplitude and phase maps of the complex residuals between the initialization and the ground truth coded surface. (c) RMSE convergence curves for ePIE, LSQ-ML, and PPTV methods. (d1-d3) Reconstructed amplitude (top) and phase (bottom) images using ePIE, LSQ-ML, and PPTV methods, respectively. Insets show magnified regions for detail comparison. Runtime and RMSE values are provided for each method. PPTV demonstrates superior reconstruction quality with the lowest RMSE and comparable runtime to ePIE and LSQ-ML.

**Experiments for biological samples imaging.** In Fig. 6, we present whole slide recovery results of various thyroid smear samples using only 50 intensity measurements, a mere fraction of the original data, reducing acquisition by a factor of ten. Given our previous analysis of the similarity between LSQ-ML and ePIE methods, differing primarily in step size adjustment, and the implications of such adjustments on reconstruction quality with limited measurements, we chose ePIE as our benchmark for comparison. In Fig. 6(a) and Fig. 6(b), PPTV reconstruction of the unstained thyroid sample outperforms ePIE, presenting noise-free imaging and revealing the internal structure of surrounding blood cells in Fig. 6(a2). Fig. 6(c) showcases PPTV reconstruction of both the amplitude and phase of the unstained thyroid smear sample, while its ePIE counterpart in Fig. 6(d) suffers from noise disturbance and loss of crucial details.

The advantages of PPTV become even more pronounced when addressing challenging scenarios characterized by pathological changes and increased tissue thickness, circumstances where traditional reconstruction methods like ePIE struggle to produce satisfactory results. To illustrate this, we selected another unstained thyroid smear sample and took 300 measurements for reconstruction, a larger dataset necessitated by the sample's complexity. As depicted in Fig. 7(a), we focus on Region-1, where lesions are present. In this area, ePIE-based recovery fails to adequately distinguish cell boundaries. In contrast, PPTV excels in recovering these tissues, presenting a line trace with better smoothness, as shown in Fig. 7(b). In Region-2, representing the recovery of normal cells, PPTV once again demonstrates its robust noise-reduction capabilities, further underscoring its effectiveness in challenging imaging scenarios.

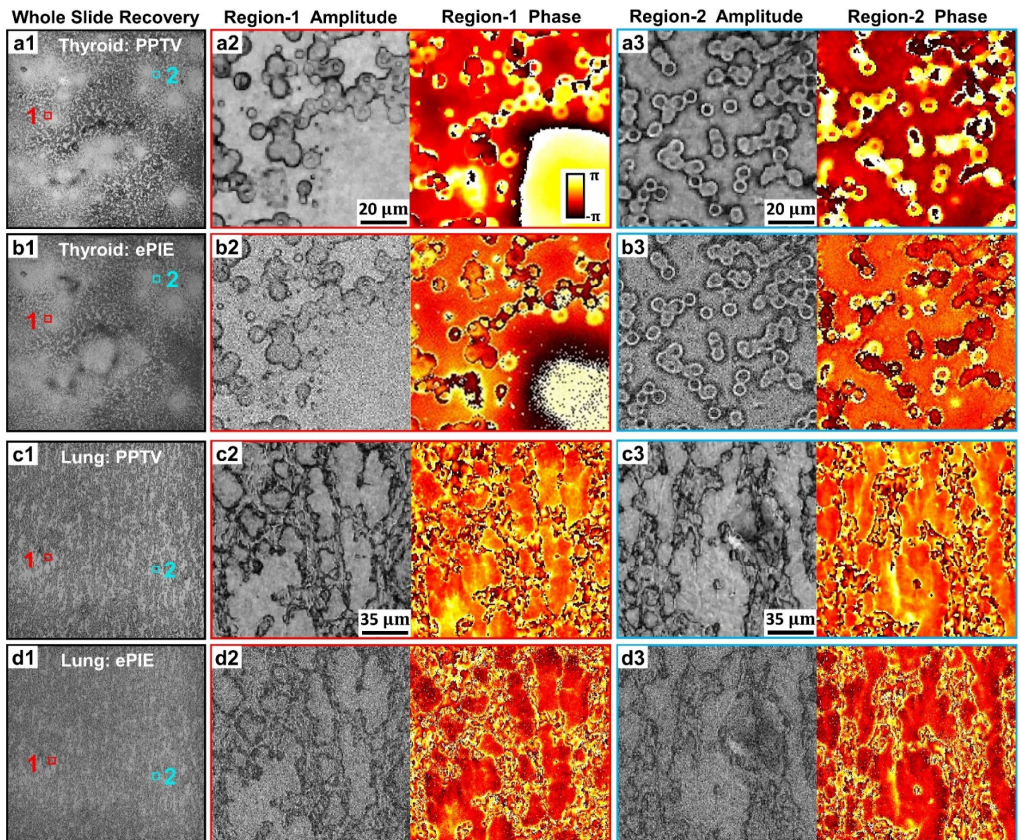

**Figure 6. Whole slide imaging of biological samples using PPTV and ePIE reconstructions.** (a1-a3) PPTV reconstruction of an unstained thyroid smear sample. (b1-b3) ePIE reconstruction of the same thyroid sample. (c1-c3) PPTV reconstruction of an unstained lung smear sample. (d1-d3) ePIE reconstruction of the same lung sample. For each row: column 1 shows the whole slide recovery; column 2 displays the amplitude and phase of Region-1; column 3 shows the amplitude and phase of Region-2.

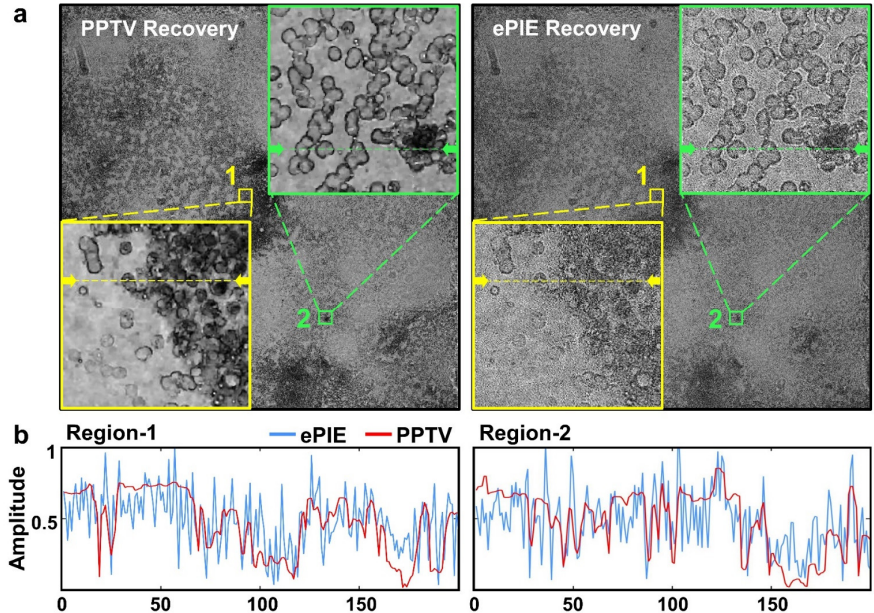

**Figure 7. Whole slide imaging of a challenging unstained thyroid smear sample.** (a) Comparison between PPTV and ePIE reconstructions. Region-1 (yellow box) shows an area with lesions, while Region-2 (green box) represents normal cells. PPTV demonstrates superior recovery of cell boundaries and internal structures in both regions. (b) Line traces of the reconstructed amplitude for Region-1 and Region-2, highlighting PPTV's ability to produce smoother, less noisy reconstructions (red line) compared to ePIE (blue line). The x-axis represents pixel position along the trace, while the y-axis shows normalized amplitude values.

These results highlight PPTV's performance in reconstructing complex biological samples, particularly in cases where traditional methods fall short. The algorithm's ability to produce high-quality images from fewer measurements, while effectively handling noise and preserving crucial details, positions it as a powerful tool for advanced ptychographic imaging in biological and medical applications.

**Discussion and conclusion**
The trade-off between measurement quantity and imaging quality has long been a significant challenge, limiting the widespread application of ptychography in optical microscopy. To address this issue, we have developed the PPTV algorithm, a novel approach that achieves high-throughput, high-resolution ptychographic microscopy while substantially reducing the number of required intensity measurements. The contributions and innovations of our reconstruction algorithm are multifaceted. We introduce an objective modeling approach for the ptychographic reconstruction problem. By incorporating sparsity regularization into the fidelity term, we fully unlock the intrinsic information contained in intensity measurements, enabling superior imaging with fewer acquisitions. Our theoretical analysis provides valuable insights into hyperparameter selection for each optimization step, offering guidance for further manual tuning and potential automated parameter adjustment. The proposed three-step optimization framework of PPTV establishes a generalized approach applicable to various ptychographic systems. It presents a practical method for implementing image priors into ptychography reconstruction through proximal algorithms.

While our work employs TV sparsity as the regularization term, it's important to note that more advanced implicit priors, such as those derived from BM3D and deep learning networks [40-45], can be seamlessly integrated into the third optimization step. This flexibility allows for future enhancements and adaptations of the algorithm to specific imaging scenarios or sample types. Although gradient sparsity has proven to be a universal prior for natural images, the quest for more suitable and effective image priors remains an essential direction for future research. Potential avenues include exploring domain-specific priors for particular types of samples, investigating multi-scale or adaptive regularization techniques, and incorporating learning-based priors that can be fine-tuned for specific applications.

In conclusion, the reported PPTV represents a step forward in alleviating the trade-off between high-quality ptychographic imaging and the resource-intensive demand for numerous measurements. By enabling high-quality reconstructions from fewer measurements, PPTV paves the way for more compact, efficient, and cost-effective ptychographic microscopy systems. This advancement has the potential to broaden the applicability of ptychography in various fields, including digital pathology, point-of-care diagnostics, endoscopy, and high-content screening, where rapid, high-resolution imaging is crucial. Future work could focus on further optimizing the algorithm's performance, exploring its application in diverse imaging scenarios, and investigating its integration with other emerging technologies in the field of computational imaging.

**Acknowledgment**
G.Z. acknowledges the support of the UConn SPARK grant. Q.Z. acknowledges the support of the UConn GE fellowship.

**Author contributions**
N.L. and G.Z. conceived the project. N.L. developed the prototype systems and prepared the display items. All authors contributed to the writing of the manuscript.

**Competing interests**
The authors declare no competing financial interest.